\documentclass[%
 reprint,
 amsmath,amssymb,
 aps,showkeys
]{revtex4-1}

\usepackage{graphics}
\usepackage{graphicx}
\usepackage{dcolumn}
\usepackage{bm}
\usepackage{hyperref}
\usepackage{placeins}
\usepackage{color}

\begin{document}

\preprint{APS/123-QED}

\title{Limitations on the number of plasmons in nanoparticles}

\author{I. A. Fyodorov}
\email{ilfedorov@gmail.com}
\affiliation{Moscow Institute of Physics and Technology, Moscow, Russia}
\affiliation{Institute for Theoretical and Applied Electromagnetics RAS, Moscow, Russia}

\author{V. M. Parfenyev}
\email{parfenius@gmail.com}
\affiliation{Moscow Institute of Physics and Technology, Moscow, Russia}
\affiliation{Landau Institute for Theoretical Physics RAS, Moscow, Russia}

\author{G. T. Tartakovsky}
\email{gtartakov@gmail.com}
\affiliation{Advanced Systems and Technologies, Irvine, CA, 92018}

\author{S. S. Vergeles}
\email{ssver@itp.ac.ru}
\affiliation{Moscow Institute of Physics and Technology, Moscow, Russia}
\affiliation{Landau Institute for Theoretical Physics RAS, Moscow, Russia}

\author{A. K. Sarychev}%
\email{sarychev$_$andrey@yahoo.com}
\affiliation{Moscow Institute of Physics and Technology, , Moscow, Russia}
\affiliation{Institute for Theoretical and Applied Electromagnetics RAS, Moscow, Russia}

\date{July 28, 2013}

\begin{abstract}
In this letter, we address the thermal processes occurring in plasmonic nanoparticles. We determine constrains imposed upon the plasmonic excitation in such nanoparticle by the resulting heat generation. Taking into account temperature dependence of the metal losses, we predict the existence of the critical number of plasmons in the nanoparticle. We also allow for the temperature dependencies of the thermal-conductivity coefficient of the environment and the Kapitza heat boundary resistance. We show that the latter can dominate the overall heat resistance in the system. Strength limitations caused by the action of electric forces are also considered. Obtained results provide instruments for the heat and strength management in the plasmonic engineering.
\end{abstract}


\maketitle

\section{\label{sec:intro}Introduction}

The modern story of the utilization of the nanoparticle-living plasmons started a decade ago, when the possibility of the enhancement of surface plasmons in the nanoparticles (NPs) by optical gain in dielectric medium was predicted theoretically \cite{Bergman2003} and demonstrated in the experiments. These results opened way to construction of the loss-free optical metamaterials \cite{Xiao2010}, subwavelength waveguides \cite{Gramotnev2010}, nanosensors \cite{Brolo2012}, and many more. After the papers \cite{Noginov2009, Oulton2009} claimed the lasing on the individual NPs, a lot of work was focused on the realization of smaller and faster sources of light \cite{Oulton2012}. A metal NP embedded in the host dielectric appeared as a unit cell of all these applications. Thermal and strength phenomena in such systems, which are surprisingly overlooked both in discussion of the experiments and theoretical considerations, are the object of the present study.

There are two main channels for the plasmonic energy decay: losses in metal and emission of photons. The former leads to the heat generation in the NP. We show, that it can readily increase its temperature up to the melting point, which is inadmissible for many applications. Another issue is the ponderomotive force, which tries to deform the NP mechanically. It is proportional to the energy density of the plasmon, which is huge due to the nanoscale mode confinement. In the following, we address these effects, considering the continuous wave (CW) and pulsed excitation of the plasmon oscillations in the NP.
\section{\label{sec:cw}Heat in CW regime}

Let us consider a spherical NP supporting a single plasmonic mode of frequency $\omega$.
We assume that the NP is permanently populated by $n$ quanta of the plasmon oscillation, which corresponds to the CW regime of the either direct or active medium-assisted excitation.
Below we will discuss heat generation only due to the decay of the plasmonic mode. Its total power, determined by the Q-factor, can be split in two parts:
\begin{equation}
\label{eq1}
\dfrac{n\hbar\omega^2}{Q}=n\hbar\omega\left(\gamma_{heat}+\gamma_{rad}\right)=P_{heat}+P_{rad},
\end{equation}
where $P$ and $\gamma$ are powers and decay rates corresponding to the losses in metal and radiation.
When the number of plasmons is fixed, heat generation is unaffected by the radiation processes. In most cases of interest, the latter is the dipole radiation, which for gold NPs in the optical range comes into play at the radius of $ \approx 20nm $ \cite{Stockman2011a}. Below this size, $ \gamma_{rad} \ll \gamma_{heat} $, and $ Q \simeq \frac{\omega}{\gamma_{heat}} $.

Let us consider the simplest experimental arrangement, when the NP is placed inside the massive bulk of the host dielectric with the heat conductivity $ \chi $. The latter removes heat from the NP according to the Fourier's law of heat conduction. In addition, we take into account the Kapitza conductance of the boundary $ h $, which allows for an abrupt temperature jump on the NP-dielectric interface, proportional to the local heat flux. It is supposed to be of critical importance on the nanoscale \cite{Jones2013}.

The temperature difference between the NP surface ($ T $) and infinity ($ T_0 $), caused by the heat generation power $ P_{heat} $ (\ref{eq1}) is given by
\begin{equation}
\label{eq2}
\delta T = T-T_0 = \dfrac{n\hbar \omega \gamma_{heat}}{4\pi a} \left( \dfrac{1}{\chi} + \dfrac{1}{ah} \right),
\end{equation}
where $ a $ is radius of the NP. Temperature inside the NP can be regarded as constant
since thermal conductivity of metal is much larger than that of dielectric host.

\begin{figure}[htbp]
\includegraphics[width=3.4in]{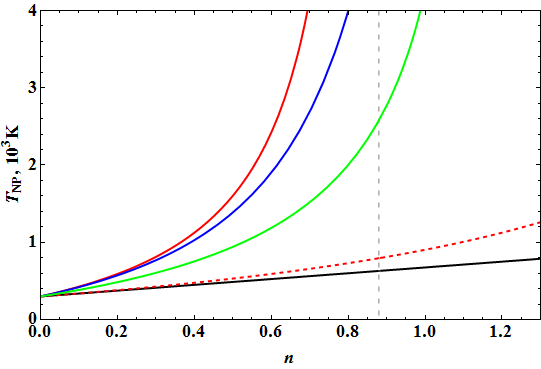}
\caption{The steady-state temperature of the NP versus the number of plasmons, red line. Red dashed line corresponds to the absence of the boundary resistance, i.e. $ h = \infty $. Black line is the most primitive estimation, where $ h = \infty $ and $ \gamma_{heat} = \gamma_{heat}(T_0) $. Vertical dashed line at $ n_{crit}\approx 0.9 $ is the critical plasmon number.
Parameters used: $a = 10nm$, $\hbar\omega = 1.9eV $, $\chi = 1 W/(m*K)$ and $ T_0 = 300^\circ K $.
Blue and green lines stand for the inclusion of temperature dependence of $ \chi $ for vitreous silica and crystalline quartz measured in \cite{Jund1999, Yoon2004}.}
\label{p1}
\end{figure}

One should account for the temperature dependence of the plasmon decay rate $\gamma_{heat}$,
due to the temperature difference $\delta T$ generally is not small.
For gold, the dependence $\gamma_{heat} (T)$ stems mainly from the electron-phonon scattering,
thus it can be approximated by linear dependence
\begin{equation}
\label{eq3}
\hbar\gamma_{heat} (T) = \alpha + \beta T.
\end{equation}
Using the data from the experimental works \cite{Liu2009a, Bouillard2012, Yeshchenko2013} one one can adopt
$\approx 1100^\circ K$ for further evaluates and assume it to be independent of frequency.
The other constant $\alpha$ has strong spread of values according to the experimental data,
we adopt $\alpha\approx 0.1$ to achieve agreement with the works
\cite{Johnson1972,Noginov2009}. Dependence (\ref{eq3}) is valid up to
$\approx 1100^\circ K $, where premelting or melting of gold starts \cite{Yeshchenko2013}
Next, we include in our calculation the temperature dependence of the boundary conductance $ h $. 
Relying on \cite{Duda2013} and assuming that the host medium is silica, 
we expect that $h\approx 5\times10^7W/(m*K^2)$ at all temperatures of the interest. 
Substituting this and (\ref{eq3}) in eq.(\ref{eq2}), we obtain temperature of the NP as a function of the plasmon population, shown in the fig.\ref{p1} as red line.

Remarkably, that temperature diverges at some \textit{critical plasmon number}
\begin{equation}
\label{eq8}
n_{crit} = \dfrac{4\pi a \chi}{\omega \beta},
\end{equation}
above which the steady state with finite temperature is impossible. The divergence point remains the same if the Kapitza effect is neglected; the corresponding result is plotted as red dashed line. Difference between solid and dashed red lines demonstrates the importance of the boundary heat resistance. Actually, the temperature jump at the interface dominates the overall temperature difference $ T-T_0 $.

Black line in the fig.\ref{p1} illustrates the most naive estimation, where $ h=\infty $ and $ \gamma_{heat} $ is taken at room temperature. As follows from the figure, is overestimates the number of plasmons needed to achieve the gold melting point up to an order of magnitude.

From (\ref{eq8}) we see, that the critical number (\ref{eq8}) is brought in by the positive value of $ \beta $ in eq.(\ref{eq3}), which is responsible for the enhancement of the electron-phonon scattering with temperature. This is expected to take place in other metals as well, though the dependence can be more complicated, as it is found for silver \cite{Yeshchenko2012}. The existence of $ n_{crit} $ (\ref{eq8}) is therefore a quite general phenomenon; its value is determined by the geometry of the nanostructure, heat-conductivity coefficients and material-dependent parameter $ \beta $.
For the gold NP with $a=10nm$ and $\hbar\omega = 1.9eV $ embedded in glass matrix with $\chi = 1 W/(m*K)$, $n_{crit} \simeq 0.9$.

Another temperature dependence which can play an essential role is that of heat conductivity of the NP environment $ \chi $. In contrast with plasmon losses and Kapitza conductance, function $ \chi(T) $ does not preserve a stable character over different common-used bulk materials. E.g., for vitreous silica it grows almost linearly between $ 100^\circ K $ and $ 1000^\circ K $ \cite{Jund1999}, while for crystalline quartz it is nearly proportional to the inverse temperature \cite{Yoon2004}. Taking into account these dependencies results in the NP temperatures plotted as blue and green lines in the fig.\ref{p1}. We see that in the first case this additional nonlinearity compensates the temperature-dependent heat generation, bringing the result back to linear type. In the second case, green line, the nonlinearities sum up, resulting in sharpening of the divergence.

The melting temperature of gold NPs with $ a=10nm $ is expected to be $ \approx 5\% $ lower than that of bulk \cite{Castro1990} and evaluates to $ T_{melt} \approx 1270^\circ K $. For the parameters from fig.\ref{p1}, $ n_{melt} \lesssim 1$. It would be extremely interesting to check this result in the experiment. $ n_{melt} $ is expected to be even less in the usual experimental setup, when NPs are placed not inside, but on the surface of the dielectric matrix \cite{Xiao2010, Oulton2009, Ma2010, Suh2012, Oulton2012}. Actually, conformational change of the NP towards the thermodynamically more stable shape is possible at substantially lower temperatures than required for the solid-liquid phase transition \cite{Link2000}. This effect depends on the initial structure of the NP, and should be analyzed in every particular case separately.

\section{\label{sec:pulse}Heat in pulsed regime}

In this section, we discuss the heat processes in the pulsed mode of plasmon excitation, in the geometry as in previous section. This regime is practically very important as a workaround for the limitations discussed in the previous section. Actually, most of the experiments on the plasmon nanolasing are performed in the pulsed mode. We are aware of only two such experiments in CW regime: \cite{Wu2011} and \cite{Lu2012}. In both of them, the full size of the resonator is approaching $ 1 $ micron, and a metal film is used as a substrate, which together makes the heat sink much more efficient than that in the system metal NP-dielectric environment.

An important timescale here is that of the electron-phonon interaction $ \tau_{ep} $. It is responsible for the equilibration of electrons and lattice temperatures and is found to be  $ \approx 3ps $, regardless of size and shape of the NP \cite{Link2000a}. Thermal dependence of metal losses (\ref{eq3}) is obtained for gold in such an equilibrium, since all the experiments we have relied on, \cite{Liu2009a, Bouillard2012, Yeshchenko2012}, are performed in the CW regime.

We start the discussion with a simple case of direct excitation of the NP by resonant light of constant power. In this case, setting time of the plasmon oscillations is $ Q/ \omega \sim 10fs $, which is the fastest timescale in the system. Number of plasmons then can be treated as a step-like function of time, determined by the intensity and duration of the pumping pulse.

\subsection{\label{sec:wpulse}Long weak pulse}

For weak pulses with duration $ \gg \tau_{ep} $, that is, starting from $ \approx 100ps $, electrons-lattice equilibrium can be assumed, and $ \gamma_{heat} $ is described by expr.(\ref{eq3}). Weakness condition assures that the temperature of the NP changes insignificantly over $ \tau_{ep} $:
\begin{equation}
\dot{T} \tau_{ep} = \dfrac{n\hbar\omega\gamma_{heat}}{C}\tau_{ep} \ll T,
\label{eq21}
\end{equation}
where $ T $ is quasi-equilibrium temperature of the NP, and $ C \approx 60 eV/K $ is its full heat capacity. For $ \hbar\omega = 1.9 eV $, $ a = 10nm $, this results in the maximum number of quanta $ n\approx 3 $.

\begin{figure}[htbp]
	\includegraphics[width=3.45in]{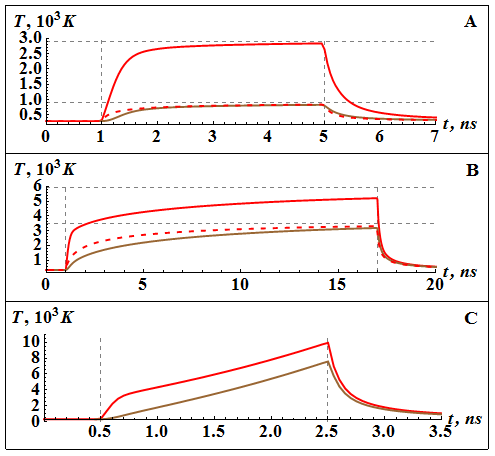}
	\caption{Dynamics of the NP temperature within a long pulse for $ n=1,2,3 $ (A,B,C). Red and brown lines are temperatures of the NP and in the dielectric near the interface. The red dashed lines show the temperature of the NP, when the boundary resistance is neglected. Vertical dashed lines indicate the exciting pulse duration.
	System parameters are as for the fig.\ref{p1}.}
	\label{p2}
\end{figure}

In this limit of long weak pulse, we obtain the transient temperature distribution using the COMSOL software package. Figure \ref{p2} shows the evolution of temperature with time for different occupation numbers $ n $. Red and brown lines stand for the temperatures at the inner and outer sides of the dielectric-NP interface.

When the plasmon number is far from the critical value, as for the case of fig.\ref{p2}A, characteristic time of the steady state establishment is $ \tau_{pp} \sim a^2/\kappa\approx 0.3ns $, where $ \kappa \approx 5\times 10^{-3}cm^2/sec $ is the glass thermal diffusivity. The steady-state temperatures agree with analytical predictions, shown as horizontal dashed lines.

As the thermal feedback activates, the energy required to bring the system to equilibrium grows faster than the plasmon number, and $ \tau_{pp} $ increases by an order of magnitude at $ n = 2 $. 
For the case $ n = 3 $, fig.\ref{p2}B, the steady state does not occur at all, in agreement with the prediction of the section \ref{sec:cw}: the critical plasmon number (\ref{eq8}) for these conditions is $ \approx 2.6 $. Instead, we observe growth of the temperature, which stops with the end of the pulse.  
The temperatures shown on figures \ref{p2}B/C will not be achieved in real experiments. Actually, the NP will disappear within $ \approx 1ns $ because of thermal damage in both cases.

Importance of the thermal boundary resistance is seen in comparison of red solid and dashed lines. Indeed, difference between the steady state temperatures is $ \approx 2\times 10^3 K $ in both fig.\ref{p2}A and fig.\ref{p2}B.


\subsection{\label{sec:spulse}Short pulse}

Now let us consider excitation of the NP by the laser pulse of duration $ \tau \sim 100fs \ll \tau_{ep} \ll \tau_{pp} $. At this timescale, electronic system of the NP is uncoupled from the metal lattice and dielectric environment, which hold their initial temperature $ T_{0} $. Energy, leaved by the pulse in the electronic system,  redistributes between the electrons and metal lattice within the next several $ \tau_{ep} $. The NP cannot melt faster than that time, provided that energy is supplied to the lattice only through the electron excitations. Actually, the melting time of the gold nanorods is found to be $ \approx 30ps $, regardless of how much energy provided by the femtosecond pulse exceeds the melting threshold \cite{Link1999b}. Next, we demonstrate some calculations, which allow us to obtain the after-pulse state of the NP.

Again, for simplicity we consider the plasmonic population of the NP as a step-like function of height $ n $. This is a good approximation for the direct excitation of plasmons, provided that $ \tau \gg 10fs $. Total energy stored in the electronic system by the end of the pulse consists of the chaotic contribution of the decayed plasmons and the coherent plasmonic oscillation:
\begin{equation}
W=n\hbar \omega \tau\gamma_{heat}(T_{0}) + n\hbar \omega.
\label{eq22}
\end{equation}

As described above, this energy goes to the lattice heating through the electron-phonon interaction. The stay-solid condition then reads
\begin{equation}
W \leqslant C(T_{melt}-T_{0}) \left[+M\right],
\label{eq23}
\end{equation}
where term in the square brackets, the full fusion heat of the NP, should be included if the condition for the full melting is needed.
\begin{figure}[htbp]
\includegraphics[width=3.5in]{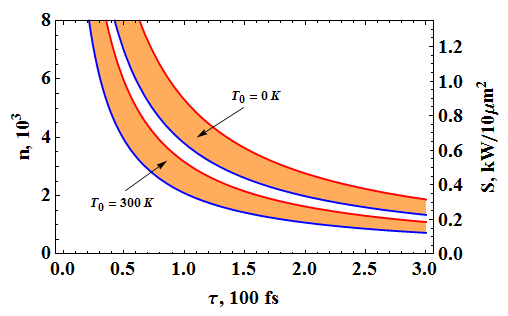}
\caption{The starting (blue) and full (red) melting conditions of the NP for its plasmon population during the pulse versus the pulse duration, for two cases of the initial temperature: $ 0^\circ K $ and $ 300^\circ K $. The filled area denotes the partial melting region. Right axis shows corresponding power of the driving laser beam. The parameters are as for the fig.\ref{p1}, with the standard thermal constants of gold \cite{Castro1990}.}
\label{p3}
\end{figure}

Fig.\ref{p3} shows the full and partial melting conditions for the NP on the $ \tau - n $ plane for two different initial temperatures $ T_{0} $. The $ 100fs $ pulse of power $ 400MW/25\mu m^2 $, which completely melted the nanorods in \cite{Link1999a}, is far in the full-melting region on the fig.\ref{p3}, as it should be.
As follows from this picture, large numbers of quanta $ \sim 10^3 $ in plasmonic systems are not forbidden in the short-pulse regime, at least due to thermal limitations.

The number of $ 1.9eV $-quanta needed for melting a gold sphere with $ a=10nm $ is $ \approx 5\times 10^4 $, so that the quantum heat features are not expected to appear at this size. It may be the case for small metal clusters of size $ \lesssim 1nm $, when $ \sim 1 $ plasmon instantaneously injected in the NP would be enough for melting. In an array of such clusters, irradiated by a laser beam, the heating and associated destruction processes would acquire a probabilistic character.

\subsection{Active excitation}

Another method of providing energy to the plasmons employs the inverted active medium coupled to the plasmonic mode of the NP. It is widely used in the experiments on the loss compensation in metamaterials and nanolasing \cite{Bergman2003, Xiao2010, Noginov2009, Oulton2009, Ma2010, Suh2012, Oulton2012}. 
In this case, a new timescale of the interaction between the active medium and the plasmonic mode, the inverse Rabi frequency, enters the dynamics of the system. 
The turn-on time of the nanolaser in the semiclassical model \cite{Stockman2011a} is found to be $ \sim 200fs $, after which the numbers of plasmons and inverted chromophores go to their stationary values. The shutdown dynamics of the nanolaser is not found in literature. Though, simple estimations with the same model show that the corresponding time is of the same order. The step-like  plasmonic population is thus a good approximation for the case of weak long pumping pulses. 

Our statement of the problem is nearly identical with that in \cite{Noginov2009}. Since the thermal damage of the nanolasers after the $ \tau=5ns $ pumping pulse is not reported, we conclude that the system is not entering the unstable regime, as demonstrated in the fig.\ref{p2}C, and a steady state occurs. 
Heat conductivity of the silica shell used in this experiment is not measured; we speculate that it is indeed lower than that of the vitreous quartz due to the porous structure of the material \cite{Blaaderen1993, Mulvaney2000}.
 
Using data for heat conductivity of vitreous silica from \cite{Jund1999}, the heat boundary resistance measured in \cite{Zhou2013a} and parameters $ a=7nm $, $ \hbar\omega=2.3eV $, $ T_0 = 290^\circ K $, $ \alpha = 0.13eV $ and $ \beta = 10^{-4}eV/K $, we find that $ n=2 $ plasmons will heat the gold core up to $ \approx 8\times 10^3 K $ and the outer surface of the silica shell up to $ \approx 2\times 10^3 K $, ensuring breakdown of the spaser.
Adopting the heat conductivity of the crystalline quartz \cite{Yoon2004} does not save the situation. We find somewhat lower temperatures up to the $ n \approx 1.35 $, where it diverges in the manner of green line on the fig.\ref{p1}.

Analysis of the experimental data leads to the conflicting estimations of the plasmon number. On the one hand, the pronounced threshold, seen in the input-output curve, is far exceeded in some of the measured points, which corresponds to the number of quanta in the laser mode is $ \gg 1 $ \cite{Rice1994}. Numerical simulation shows that this situation would result in the fast increase of the temperature far above the melting and boiling points of gold and silica, which certainly should destroy the nanolaser. 
On the other hand, the total pumping power $ W $ absorbed per one nanolaser, which is calculated in \cite{Noginov2009}, is connected with the number of plasmons as
\begin{equation}
W = n\hbar \omega\left(\dfrac{\omega\tau}{Q}+1\right).
\label{eq24}
\end{equation}

Allowing for the temperature dependence of the $ Q $-factor as before, near the threshold ($ W = 10^{-13}J $) one obtains $ n\approx 0.15 $ and the core temperature $ T\approx 10^{3} K $ both for vitreous and crystalline quartz heat conductance of the shell. 
The most intense pumping, $ W = 5\times 10^{-13}J $, leads to $ n\approx 0.5 $, $ T\approx 2\times 10^{3} K $, while temperature of the outer shell surface is $ \approx 570^\circ K $ in case of vitreous silica and $ \approx 310^\circ K $ for the crystalline quartz. In both cases, it stays solid and damage of the nanolaser is not expected.

We think that inconsistency in the plasmon number estimations arises from the assumption that the plasmonic mode is localized on a single nanoparticle. In our opinion, the observed phenomenon can be accounted for a collective effect resembling those in random lasers \cite{Wiersma2008}. The measured many-quanta laser mode is then distributed over a large number of the core-shell NPs with individual plasmon populations $ \sim 0.1 $. This expounding is consistent with thermal considerations discussed above, but arises a new question about the results obtained with a diluted sample. Actually, the lasing mode could survive the diluting if it is localized in a set of nanoparticles. 

The regime of short pulse takes place in experiments \cite{Oulton2009, Ma2010, Suh2012}, where the pumping duration is $ 45-100fs $. It is of order of the equilibration times in the nanolaser, and the results of the section \ref{sec:spulse} cannot be used to quantitatively predict heating in the system. Though, with the $ n $ in the second summand in (\ref{eq22}) replaced by the $ n + number~of~excited~emitters $, the condition (\ref{eq23}) remains valid. Accurate treatment of the fast heat processes in the active plasmonic devices needs a detailed analysis, including the electron-lattice energy exchange kinetics.
\section{\label{sec:force}Ponderomotive forces}

Another kind of limitations imposed on the operation of plasmonic devices arise from the  ponderomotive forces \cite{Landau1984}.
Let us consider a high-Q spherical nanoparticle of radius $a$ in thermal and mechanical equilibrium. Due to the quasistatic approximation one can neglect the Abraham force and the expression for the bulk density of the electric part of ponderomotive forces then reads ${\bm f} = \nabla \hat\sigma$, 
where the stress tensor
\begin{equation}
\label{eq10}
\hat \sigma = \frac{(2\varepsilon\overline{{\bf E}{\bf E}^{\scriptscriptstyle T}} - \overline {E^2}\hat I)}{8\pi},
\end{equation}
with $ \textbf{E} $ is the electric field,
overline stands for time averaging and $\varepsilon$ is the dielectric permittivity,
subscript `T' denotes transposition and $\hat I$ is identity matrix.
Deriving (\ref{eq10}) we adopted the simplest model, when the dielectric susceptibility $\varepsilon-1$ 
is proportional to total density of the media and there is no influence on the susceptibility 
from the deformations preserving the density.

\begin{figure}[t]
	\includegraphics[width=1.45in]{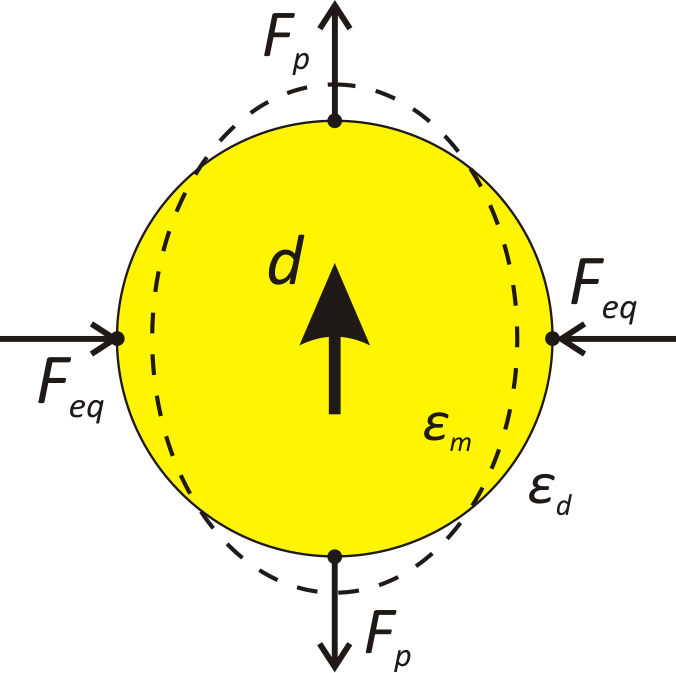}
	\caption{The ponderomotive forces in the case of spherical NP with dipole mode. They tend to deform the sphere into an ellipsoid, elongated along the direction of the dipole moment.}
	\label{p5}
\end{figure}

In the case of NP with the most-used dipole mode, the ponderomotive forces per unit area acting on the boundary NP-dielectric enviroment at equator and poles are given by
\begin{equation}
\label{eq11}
F_{eq} = (\varepsilon_d+\varepsilon_m-2\varepsilon_d\varepsilon_m)F,
\qquad
F_p =  (\varepsilon_d+\varepsilon_m)F,
\end{equation}
where the typical force $F=(\varepsilon_d - \varepsilon_m)\overline{\textbf{E}_0^2}/(8\pi\varepsilon_d^2)>0$ 
and $\textbf{E}_0>0$  is the electric field inside the nanoparticle, $\varepsilon_m$ and $\varepsilon_d$ are the permittivities of NP and dielectric medium respectively. The direction of forces represented by the arrows in fig.\ref{p5}. Note, that in general dielectric permittivity is complex quantity $\varepsilon = \varepsilon' + i \varepsilon''$, but in case of high-Q nanoparticle one should assume $\varepsilon \approx \varepsilon'$.

The electric field $\textbf{E}_0$ can be expressed through the number of plasmons $n$ and Q-factor
\begin{equation}
\label{eq12}
\dfrac{\omega}{4 \pi} \varepsilon_m'' \overline{\textbf{E}_0^2} a^3  = \dfrac{n \hbar \omega^2}{Q},
\end{equation}
Substituting this into (\ref{eq11}) and using the relation $Q =- \varepsilon_m' / \varepsilon_m''$ \cite{Stockman2011a}, we obtain
\begin{equation}
\label{eq13}
F_{eq} = \dfrac{n\hbar\omega}{2a^3} \left( 1 - \dfrac{\varepsilon_d}{\varepsilon_m} \right).
\end{equation}

The resonance condition on sperical nanopartical is $2 \varepsilon_d + \varepsilon_m = 0$.
Setting $ \hbar\omega = 1.9 eV $ and $ a=10nm $, one gets $F_{eq} \approx n\times 2.3\times 10^5 Pa$. For $ n=500 $ plasmons, this value of order of the initial yield stress of the nano-sized gold \cite{Lee2007}. In addition, the elastic properties of gold are known to depend strongly on temperature \cite{Condra1975}, so that electro-mechanical deformations of plasmonic nano-structures can be expected in the short-pulse regime of plasmon excitations.

\textcolor{black}{Deformation of the NP in turn affects the resonance frequency of the plasmonic mode. In case of sphere-to-ellipsoid deformation, the dipole resonance frequency splits into the longitudinal and transversal doublet. Splitting magnitude exceeds the plasmon linewidth already at ellipsoid axes relation of about $ 1.2 $. This effect provides a possible surviving path of the NP against the heating death. On the other side of the coin, ponderomotive forces can lead to the fragmentation of the NP. Experimental evidence of this phenomenon is given in ref.\cite{Link1999a} for the nanorod geometry.}



\section{Conclusion}

Temperature dependence of the plasmon decay rate in metal is shown to play a principal role in the heat processes occuring in the plasmonic nanosystems. Resulting nonlinearities lead to the emergence of the critical plasmon number, which can be of order of unity for real experimental setups. Beyond this number, the stable heat solution does not exist. In this situation, behavior of the plasmon in the NP is essentially non-classical.

Still, the pulsed operation regime allows one to populate the NP with large numbers of plasmons on the femtosecond timescale. On the other hand, population of the plasmons in the NP is restricted from above by the ponderomotive forces. For the realistic parameters, these forces can be large enough to cause mechanical damages of the nanosystem.

There is still lack of the experimental data for the thermal and strength phenomena in plasmonics, which could verify predictions made in this paper. A systematic study of these effects would be of great help for understanding the challenges, which will be faced on the way to real-world applications.

\FloatBarrier
\bibliography{library}

\end{document}